\documentstyle[12pt]{article}

\def\be{\begin{equation}}
\def\ee{\end{equation}}
\def\bea{\begin{eqnarray}}
\def\eea{\end{eqnarray}}
\def\Ja{J_\mu(\phi)}
\def\Jb{J^\mu(\phi)}
\def\l{l_0}
\def\Na{N_\mu}
\def\Nb{N^\mu}

\def\pa{\partial_\mu}

\def\pb{\partial^\mu}
\def\pd{\Box}

\begin{document}
\pagestyle{plain}
\begin{center}
\LARGE{\bf Vacuum-Structure and a Relativistic Pilot Wave\\}
\vspace{.5cm}
\small
\vspace{1.5cm}
{\Large{\bf H. Salehi$^{1,2}$}}\footnote{e-mail address: h-salehi@cc.sbu.ac.ir},~~
{\Large{\bf H. Motavali$^{3}$}}\footnote{e-mail address: motaval@theory.ipm.ac.ir}
and {\Large{\bf M. Golshani$^{4}$}}\footnote{e-mail address: golshani@ihcs.ac.ir}  \\  \vspace{0.5cm}
\small$^{1}$ Department of Physics, Shahid Beheshti University, Evin, Tehran 19834,  Iran \\
$^{2}$ Institute for Humanities and  Cultural Studies, 14155-6419, Tehran, Iran\\
{$^{3}$ Department of Physics, Amirkabir University of Technology, 15875-4413, Tehran, Iran \\
$^{4}$ Department of Physics, Sharif University of Technology, 11365-9161, Tehran, Iran}\\
\today
\end{center}
\vspace{.5cm}
\small
\date{\today}
\begin{abstract}
We study a model for analyzing the effect of a principal violation of the
Lorentz-invariance on the structure of vacuum. The model is based
on the divergence theory developed by Salehi (1997). It is shown that the
divergence theory can be used to model an ensemble of particles.
The ensemble is characterized by the condition that its members are basically
at rest in the rest frame of a preferred inertial observer in vacuum.
In this way we find a direct dynamical interplay between a particle and its
associated ensemble. We show that this effect can be understood in terms
of the interaction of a particle with a relativistic pilot wave through an
associated quantum potential.
\end{abstract}
\section{Introduction}
The requirement of the Lorentz-invariance is one of the basic foundations of modern
physical theories. This principle states that all inertial frames of reference
are equivalent. A key feature of the Lorentz-invariance would be its
violation. Typically, such a violation would single out an individual
frame in which a preferred inertial observer is at rest. The four-velocity of
this preferred inertial observer
may be interpreted as a time-like vector field which
has almost the same value
throughout the Einstein-Minkowski space. Such an absolute
object in vacuum, which is an example of what is usually known as
the internal vector $N_\mu$ (Blokhintsev, 1964; Nielsen and Picek, 1983; Phillips, 1965)
may generally be considered to constitute one of the basic characteristics
of the violation of Lorentz-invariance. \\
Another physical characteristic of
a Lorentz-noninvariant vacuum is provided by
the concept of the universal length $l_0$
(Blokhintsev, 1964; Nielsen and Picek, 1983) for which we adopt the following interpretation:
The length $l_o$ is used to define
a minimal limiting length for all physical
distances probed in 'typical' measurements. By a typical measurement
we mean one which can be performed relative to a
large (ideally infinite) number of physically equivalent inertial frames.
Clearly, the length $l_0$ acts as an
absolute demarcation line between macroscopic (large)
distances and microscopic (small) ones. It
refers to an absolute property of a
Lorentz-noninvariant vacuum. \\
Such possible limitations on the requirement of the exact
Lorentz-invariance
in vacuum, as reflected in
the universal length $l_0$ and the associated internal
vector $N_{\mu}$, lead to immediate limitations on the applicability of the
standard form of
relativistic motion of a free particle.
In fact, if the Compton wave-length (the typical size) of
a free particle in vacuum
approaches the universal length $l_0$ we expect that the
number of real systems which could act as equivalent
inertial systems characteristic
of the
relativistic motion of the particle would reduce drastically. Specifically,
we expect that the characteristic rest frame of
a free particle with a Compton wave
length $\sim l_0$ to correspond to the rest frame of
the preferred inertial observer characterized by the internal
vector $N_{\mu}$. Thus, for such a particle a strong alteration of the
standard form of
the relativistic motion predicted by special relativity could in
principle arise. The purpose of this paper is a consideration of this issue.\\
The organization of the paper
is as follows. Section 2 provides a formal scheme for
incorporating the invariance breaking effect of the universal length and the
associated internal vector into the structure of vacuum in Einstein-Minkowski
space. This scheme is presented in terms of the divergence theory developed by
Salehi (1997). The divergence theory admits the treatment of
some general constraints imposed
on the vacuum by the condition of Lorentz-noninvariance. Also, we shall
impose a boundary condition on the configuration of vacuum by combining
the divergence theory with a time-asymmetric law discussed by Salehi and Sepangi (1999).
Section 3 deals with the
particle interpretation of the theory. The particles associated with the
divergence theory may be interpreted as an ensemble of free particles with
Compton wave-lengths having the order of magnitude of
the universal length. This ensemble arises
as a particular solution of the divergence theory and is found to be
characterized by the condition that its rest frame corresponds to the
rest frame of the preferred
inertial observer characterized by the internal vector. It appears that the type of the
particle motion within the ensemble is strongly constrained by this condition.
Actually we find a dynamical interplay between an individual particle and its
associated ensemble through a quantum potential generated by a
pilot wave which we shall finally derive in the section 4.
\section{Divergence theory}
In order to study the violation of Lorentz-invariance in vacuum,
the basic idea is
to consider
a dynamical coupling of a real scalar field $\phi$ with the
internal vector $N_\mu$ in Einstein-Minkowski space.
Following (Salehi, 1997) we start with the consideration of the current
\begin{eqnarray}
{\Ja}=-\frac{1}{2}{\phi}\stackrel{\leftrightarrow}{\partial_\mu}
\phi^{-1}.
\end{eqnarray}
It is easy to show that
\begin{eqnarray}
{\Ja}{\Jb}=\phi^{-2}{\pa {\phi}\pb {\phi}}
\end{eqnarray}
and
\begin{eqnarray}
{\pa \Jb}=\phi^{-1}({\Box {\phi}}-\phi^{-1}\pa {\phi}\pb {\phi})
\end{eqnarray}
in which, $\Box=\eta_{\mu\nu} \partial^\mu \partial^\nu$ denotes the d'Alembertian associated with the Minkowski-metric $\eta_{\mu\nu}$.
Combining the relations (2) and (3) we get the identity
\begin{eqnarray}
{\Box {\phi}}-[{\Ja}{\Jb}+\pa \Jb] {\phi}=0.
\end{eqnarray}
It should be noted that, this identity is a consequence of the
definition (1) and not a dynamical law for $\phi$. However, one can obtain
several dynamical theories by making various assumptions about
the source of the divergence $\pa \Jb$ in (4).

We analyze a model in which the source of the divergence is specified as
(Salehi and Sepangi, 1999)
\begin{eqnarray}
{\pa {\Jb}}={\l^{-1}}{\Na}{\Jb}.
\end{eqnarray}
The corresponding field equation is
\begin{eqnarray}
\Box \phi-[\Ja \Jb+\l^{-1} \Na \Jb]\phi=0.
\end{eqnarray}
Under the duality transformation
\begin{eqnarray}
(\phi,\Na,\eta_{\mu\nu}) \rightarrow (\phi^{-1},-\Na,\bar{\eta}_{\mu\nu})
\end{eqnarray}
in which ${\bar{\eta}}_{\mu\nu}$ is given by
\begin{eqnarray}
{\bar{\eta}}_{\mu \nu}=\eta_{\mu \nu}+2\Na N_\nu
\nonumber
\end{eqnarray}
equation (6) transforms to
\begin{eqnarray}
\Box \phi-[\Ja \Jb+\l^{-1} \Na \Jb-2 N_\mu N_\nu \partial^\nu \Jb]\phi=0.
\nonumber
\end{eqnarray}
This equation can be rewritten as
\begin{eqnarray}
\Box \phi-[\Ja \Jb+\pa {\Jb} -2\l \frac{d}{d \tau}(\partial_\mu \Jb)]\phi=0
\end{eqnarray}
where we have used
\begin{eqnarray}
N_\nu \partial^\nu=\frac{d}{d \tau}
\nonumber
\end{eqnarray}
which is the defining relation of what we call the internal time-parameter $\tau$
associated with $N_\mu$.\\
Comparison of the last two terms in the brackets in equation (8) indicates that the physical effect
of the last term is smaller than its previous term as a result of smallness
of the universal length $\l$.
Therefore, the last term may be neglected \footnote{This assumption has been
implicitly considered by Salehi and Sepangi (1999).} and equation (8) reduces to equation (6).
As a result, equation (6) becomes invariant under the duality transformation (7).
The meaning of this duality is that, if we start from the configuration
$(\phi,\Na,\eta_{\mu\nu})$ on a Lorentzian domain (a domain in Einstein-Minkowski space
with the metric $\eta_{\mu \nu}$ and
signature $(-+++)$) we get another
configuration by the duality transformation, namely $(\phi^{-1},-\Na,\bar{\eta}_{\mu\nu})$,
on an Euclidean domain (with the metric ${\bar{\eta}}_{\mu \nu}$ and
signature $(++++)$). Therefore the duality transformation connects equivalent
configurations of vacuum with different signatures. \\
To fix the causal structure of the Einstein-Minkowski space for our analysis, it is
necessary to combine equation (5) with a duality breaking condition. On a
Lorentzian domain we adopt the time-asymmetric condition (Salehi and Sepangi, 1999)
\begin{eqnarray}
{\pa {\Jb}}={\l^{-1}}{\Na}{\Jb}>0.
\end{eqnarray}
Clearly this condition can not simultaneously be satisfied for configurations
related
by the duality transformation, because the source of the divergence in (9)
will reverse the sign under that transformation.
Therefore, the condition (9) implies an assertion about both a preferred arrow
of time, namely that determined by the internal vector $N_\mu$, and a preferred
configuration of the scalar field $\phi$.

Now we proceed to derive a general constraint imposed on the configuration of the
scalar field $\phi$ by the condition of the duality breaking.
If one combines the divergence law (5) with (1), one finds
\begin{eqnarray}
\pa ({\Jb}-{\l^{-1}}{\Nb}{\ln\phi})=0.
\nonumber
\end{eqnarray}
We focus here on a particularly simple solution of this equation, given by
the time-like current
\begin{eqnarray}
{\Ja={\l^{-1}}{\Na}{\ln\phi}}.
\end{eqnarray}
Using this solution, we may then rewrite the divergence law (5) as
\begin{eqnarray}
{\pa {\Jb}}=-{\l^{-2}}{\ln\phi}.
\nonumber
\end{eqnarray}
Now, to conform with the duality breaking condition (9), one
should require
\begin{eqnarray}
\phi(x)<1
\end{eqnarray}
which constraints the configuration of the scalar field $\phi$.

\section{The relativistic ensemble of particles}
The divergence theory (5) was thought as a method for studying
the effect of the Lorentz-noninvariance on the structure of the vacuum.
Now, our purpose is to demonstrate that it involves certain features that
can be interpreted in terms of an associated ensemble of particles. \\
The key idea is to take the current $J_{\mu} (\phi)$ as the defining characteristic
of a fictitious ensemble. In specific terms, a particle in the
ensemble that may potentially pass through a given point is assumed to have
a four-momentum characterized by $J_{\mu} (\phi)$. Because of (10), such a particle
would be absolutely at rest in the rest frame of the preferred inertial
observer characterized by the internal vector. \\ It is possible to convert
the quantitative description of the ensemble into a more standard from.
In fact, combining (6) with (1) and (10), we can derive
\begin{eqnarray}
\partial_{\mu}S\partial^{\mu}S={\l^{-2}}{\ln\phi}+(\frac{\pd {\phi}}{\phi})
\end{eqnarray}
with $S=\ln\phi+\alpha$ where $\alpha$ is a constant.
All terms in this equation are form-invariant
under the scale transformation $\phi\rightarrow \lambda\phi$,
$\lambda$ being
an arbitrary constant, except
the first term on the right hand side. Thus, this term is sensitive
to the background average value $\bar{\phi}$ of the scalar field $\phi$.
We may, therefore, use a background approximation for this term, namely
by replacing
the $\ln\phi$ by $\ln{\bar{\phi}}$ which should be negative due to (11). We shall
subject the choice of $\bar{\phi}$ to the condition that $\ln\bar{\phi}$ shall adjust
${\l^{-2}}\ln\bar{\phi}$ to the characteristic mass-scale ${\l^{-1}}$
of the theory, leading to
\begin{eqnarray}
\partial_{\mu}S\partial^{\mu}S=-M^{2}+(\frac{\pd {\phi}}{\phi})
\end{eqnarray}
with $M^{2}=-{\l^{-2}}\ln{\bar{\phi}}\approx l_{0}^{-2}$.
Equation (13) can be interpreted as the
Hamilton-Jacobi equation characterizing the ensemble.
In this ensemble the particle-mass M
is adjusted to ${l_{0}}^{-1}$ by
a corresponding background average value of the scalar field. Furthermore, the
term $\frac{\pd {\phi}}{\phi}$ indicates a modification of the particle mass
due to a local deviation of the scalar field from its background average value.
This term reflects a direct interplay between a particle
and its associated ensemble.

We should emphasize the distinguishing feature of equation (13).
The dynamical properties of the particles in the ensemble
(the configuration of the S-function up to an additive constant)
basically emerges as a consequence of a local deviation of the scalar field
from its background average value $\bar{\phi}$. Moreover, the latter value
serves to adjust the mass of the particle in the ensemble to the characteristic
mass-scale of the theory, namely ${l_{0}}^{-1}$. In this way
the scalar field $\phi$ is found to carry the whole dynamical
characteristics of an ensemble of particles.\\  Now, we try to characterize
the ensemble in terms of a hydrodynamical equation. To derive such an equation
we begin with
\begin{eqnarray}
\pa({\phi^2\Jb})&=&
2\phi\pa\phi\Jb+\phi^2 l_0^{-1} N_\mu \Jb       \nonumber\\
                  &=&2\phi^2 \Ja\Jb+\phi^2 l_0^{-1}N_\mu \Jb.  \nonumber
\end{eqnarray}
To rewrite this equation in a convenient form, we linearize the quadratic
term in $J_{\mu}(\phi)$ on the right hand side to obtain the approximate relation
\begin{eqnarray}
\pa({\phi^2\Jb})\approx 2\phi^2 \overline{J}_\mu(\phi)\Jb+\phi^2 l_0^{-1} N_\mu \Jb.
\nonumber
\end{eqnarray}
Here the constant current $\overline{J}_\mu (\phi)$ may be obtained from (10) by
replacing $\phi$ by its corresponding
background average value of $\bar{\phi}$, used previously
in the derivation of equation (13) from equation (12). Actually we get
$\overline{J}_\mu (\phi) \approx -l_{0}^{-1}N_{\mu}$, leading to
\begin{equation}
\pa({\phi^2\Jb})\approx -\phi^2 l_0^{-1}N_\mu \Jb.
\end{equation}
Using the definition (1), equation (14) can be brought into the form
\begin{eqnarray}
{\pa({\phi^2\pb S})}\approx -\phi^2 \l^{-1} \frac{d }{d \tau} \ln\phi.
\end{eqnarray}
The right hand side of this equation, which in principle may be
different from zero, indicates
that, if we relate $\phi^{2}$ to the particle-number density, the
conservation of particles in the ensemble may be violated.

\section{Derivation of the pilot wave }
It is possible to interpret the equation of motion for a particle  in the
ensemble in terms of a pilot wave. Consider the wave
\begin{eqnarray}
\psi= \phi~ e^{iS}
\nonumber
\end{eqnarray}
which satisfies, as a consequence of equations (13) and (15), the equation
\begin{eqnarray}
\pd {\psi}-{M^2}{\psi}\approx -({i}{\l}^{-1} \frac{d}{d \tau}\ln|{\psi}| )\psi.
\nonumber
\end{eqnarray}
This equation reduces to the relativistic Klein-Gordon equation in the idealized
limit in which any conceivable dependence of $\psi$ on the internal
time-parameter $\tau$ is ignored.\\
The merit of introducing the wave $\psi$ is that it acts
as a sort of a pilot wave in the sense of causal interpretation of
quantum mechanics (Holland, 1993; Bohm, 1952; Bohm and Hiley, 1993), the term
$\frac{\pd {\phi}}{{\phi}} $ on the right
hand side of equation (12) being the associated quantum potential. \\
It should be emphasized that the pilot wave derived in this section is thought
to mediate merely the interaction of a particle with its associated ensemble.
The necessity of this interaction is deeply connected with the fact that the
characteristic rest frame of the ensemble at every point appears to be
just the rest frame of the preferred inertial observer
characterized by the internal vector, as reflected in (10). The
corresponding quantum potential just serves to incorporate this common global
memory effect. We should note that this behavior does not seem to
be symptomatic of any relativistic particle. In fact, if the Compton
wave-length of a relativistic particle is much larger than the universal length
$l_0$, we generally expect to find a large number of real systems as
equivalent inertial systems characteristic of the relativistic motion.
However, this number may reduce drastically as the mass of a relativistic
particle approaches the mass-scale $l_0^{-1}$. One may ask how this expected
universal scaling behavior common to all relativistic particles is related to
their quantum behavior. The exploration of this question is an interesting
subject which requires considerable clarification.\\

\end{document}